# PRACTICAL RANGE AGGREGATION, SELECTION AND SET MAINTENANCE TECHNIQUES

Mugurel Ionuţ ANDREICA[1], Nicolae ŢĂPUŞ[2]

*În acest articol prezentăm câteva metode şi tehnici noi, foarte practice, pentru calculul unor valori agregate în structuri şi baze de date multidimensionale. De asemenea, considerăm şi problema determinării eficiente a celei de-a k-a valori minime din mulţimi definite prin diverse constrângeri. A treia contribuţie a articolului este reprezentată de câteva extensii şi aplicaţii ale unor probleme fundamentale de gestiune a mulţimilor de elemente.*

*In this paper we present several new and very practical methods and techniques for range aggregation and selection problems in multidimensional data structures and other types of sets of values. We also present some new extensions and applications for some fundamental set maintenance problems.*

## 1. Introduction

In this paper we present several novel techniques for range aggregation, selection and the maintenance of sets of elements under certain constraints, as well as some extensions and applications. Range aggregation and selection are two fundamental problems with applications in a wide range of domains. *Range aggregation* techniques were developed in the context of OLAP data cubes [1], computational geometry problems and data structures [2], (multidimensional) databases [3, 4], and so on. The *range selection* problem considers the computation of the $k^{th}$ smallest value among a set of (database) entries whose attributes belong to a given range. Selection techniques have been developed in the context of geometric inter-distances [9] and implicit sorted matrices [5]. *Set maintenance* is another fundamental issue with applications in a wide range of domains. The rest of this paper is structured as follows. In Section 2 we consider some range aggregation problems, while in Section 3 we discuss several selection and range optimization problems. Sections 2 and 3 also address several set

[1] Assist., Computer Science and Engineering Department, Faculty of Automatic Control and Computer Science, Politehnica University of Bucharest, Bucharest, Romania, email: mugurel.andreica@cs.pub.ro
[2] Prof. Dr., Computer Science and Engineering Department, Faculty of Automatic Control and Computer Science, Politehnica University of Bucharest, Bucharest, Romania, email: nicolae.tapus@cs.pub.ro

maintenance issues. In Section 4 we conclude.

## 2. Range Aggregation Problems

In this section we consider several range aggregation problems with applications in a wide range of fields. The general model is the following. We have *n* points in a d-dimensional space. Each point *i* has *d* coordinates *(x(i,1), …, x(i,d))* and a weight *w(i)*. We are interested in answering efficiently the following types of queries : compute an aggregate of the weights of all the points in a given d-dimensional range *[xa(1),xb(1)] x … x [xa(d),xb(d)]*. We will consider two cases : the *sparse* case and the *dense* case. Let's consider that the points have *m(j)≤n* distinct coordinates in dimension *j*. When *m(j)* is significantly smaller than *n (1≤j≤n)*, we call the point set *dense*. For instance, if $m(*)=O(n^{1/d})$, the points are densely packed into a d-dimensional "cube" of side lengths *m(*)*.

We will start with the sparse case, for which the multidimensional range tree data structure [2] is probably the best known. A d-dimensional range tree consists of a balanced tree constructed on the *m(d)* distinct $d^{th}$ coordinates of the *n* points. The values of the $d^{th}$ coordinate are the keys of this balanced tree. A special property of this tree is that the keys are stored only at its leaves. An inner node *q* of the tree contains the smallest coordinate *left(q)* and the largest coordinate *right(q)* of a leaf in the subtree of *q* (for a leaf, we consider *left(q)=right(q)*). Each tree node *q* (leaf or inner node) contains a (d-1)-dimensional range tree *T(q)*, constructed only over the points whose $d^{th}$ coordinate is in the range *[left(q),right(q)]*. When *d=1*, instead of *T(q)*, every node *q* stores an aggregate value *qagg*, representing the value of the aggregate function over the weights of the points contained in node *q*'s subtree. For *d=1*, the *qagg* value of a leaf node is the weight of the point corresponding to that leaf (if there are several points with exactly the same coordinates, we replace them by a single point whose weight is the aggregate of the weights of the points). For *d=1* and an inner node *q*, we have *qagg(q)=aggf(leftson(q), rightson(q))*, where *aggf* is the aggregate function and *leftson(q)* and *rightson(q)* denote the left and right sons of node *q*. A range aggregate query on a d-dimensional range *[xa(1),xb(1)] x … x [xa(d), xb(d)]* is performed by computing in *O(log(n))* time a *canonical decomposition* of *O(log(n))* tree nodes *q* in the $d^{th}$ dimension, such that their *[left(q), right(q)]* intervals are disjoint and, together, these intervals contain all the distinct coordinates contained in *[xa(d),xb(d)]* (and only these). Then, a (d-1)-dimensional range aggregate query is performed on the *T(q)* trees of each node *q* in the canonical decomposition and the results are aggregated. When *d=1*, the *qagg* values of the nodes in the canonical decomposition are aggregated. Thus, a query takes $O(log^d(n))$ time. The range tree can be turned into a dynamic data structure which supports insertions of new points and deletions of old points. In this paper we are concerned only with a semi-dynamic version of the range tree, in which the weights of the points can be changed, but the points themselves cannot be deleted (nor can new points be inserted). Note, though, that a logical deletion can be performed, by setting the weight of a point in such a way that it does not influence the values of the aggregate function (e.g. to $-\infty$ for *aggf=max*, *0* for *aggf=+*, *1* for *aggf=\**, or $+\infty$ for *aggf=min*). If we want to change the weight of a point, we find the *O(log(n))* nodes *q* in the d-dimensional tree, such that the $d^{th}$ coordinate of the point is in the range *[left(q),right(q)]* ; then, we call the update function on the (d-1)-dimensional trees *T(q)*, considering only the first *(d-1)* coordinates of the point. When *d=1*, after locating the leaf containing the point, we change the point's weight and recompute the *qagg* value of the leaf. Then, we recompute the *qagg* values of each of the leaf's ancestors, from the leaf towards the root, by considering the *qagg* values of their left and right sons. A weight update takes $O(log^d(n))$ time. We can extend the range tree with a range update function : *rangeUpdate(u, [xa(1),xb(1)] x … x [xa(d),xb(d)])*. This function has the effect of setting the weight of each point *i* in the range to *uaggf(u, w(i))* (*uaggf* is the update function). We can implement this function efficiently by using the ideas presented in [7], where the authors described an algorithmic framework based on segment trees (see Fig. 1) in the *dense* case. Each node *q* will maintain two trees, $T_1(q)$ and $T_2(q)$. One of them will be updated whenever *q* is part of the canonical decomposition of a range given as argument to *rangeUpdate*. The other one will be updated whenever *q* is the ancestor of a node *p* which is part of the canonical decomposition of a range. When *d=1*, each tree node *q* maintains an extra value *uagg*, representing the aggregate of all the update values *u* of the range update calls for which node *q* was part of the canonical decomposition of the update range (in the first dimension). The exact details can be derived from the case presented in [7] (note that if multiple points with the same $1^{st}$ coordinate were replaced by a single point, we may also need to maintain how many of the original points are represented by the stored point). We will now consider the dense case. We focus here only on range aggregates where the aggregate function *aggf* is invertible, i.e. if *c=aggf(a,b)*, then *a=aggf(c,b$^{-1}$)* (e.g. *aggf=addition* and *(aggf)$^{-1}$=subtraction* ; *aggf=multiplication* and *(aggf)$^{-1}$= division* ; *aggf=xor* and *(aggf)$^{-1}$ =xor*, and so on). Moreover, we will be interested only in static data sets at first. We can think of the points as being located in a d-dimensional "cube" of size *m(1) x m(2) x … x m(d)*. Each cell *(c(1), …, c(d))* of the cube *(1$\leq$c(i)$\leq$m(i)* ; *c(i)* stands for the $c(i)^{th}$ distinct coordinate value in dimension *i* , *1$\leq$i$\leq$d*) is occupied by a point *i* and we say that *Cube(c(1), …, c(d))=w(i)*. If there is no point with the corresponding coordinates, we say that *Cube(c(1),…,c(d))* is equal to the neutral element of the aggregation function (e.g. *0* for *+,xor* ; *1* for *\**, and so on). We want to compute the aggregate of the cube values in a d-dimensional range *[clow(1), chigh(1)] x … x [clow(d), chigh(d)]*, where $1 \leq clow(i) \leq chigh(i) \leq m(i)$ ($1 \leq i \leq d$). We will present a well-known technique, based on computing the prefix aggregate cube *PSCube*, where *PSCube(c(1), …, c(d))* is the aggregate value of all the entries *Cube(c'(1), …, c'(d))*, with $1 \leq c'(i) \leq c(i)$ ($1 \leq i \leq d$). We will first assume that this prefix cube is already computed and we will show how we can use it. Later, we will show how to compute the prefix cube, too. Let's assume that we want to compute the aggregate of all the entries in a range *[clow(1), chigh(1)] x … x [clow(d), chigh(d)]*. We can do this by aggregating $2^d$ entries of the prefix cube *PSCube* :

$$RangeQuery(\prod_{i=1}^{d}[clow(i), chigh(i)]) = \sum_{\substack{s(i) \in \{clow(i)-1, chigh(i)\} \\ 1 \leq i \leq d}} (-1)^{\sum_{j=1}^{d} parity(s(j))} \cdot PSCube(s(1),...,s(d)) \cdot \quad (1)$$

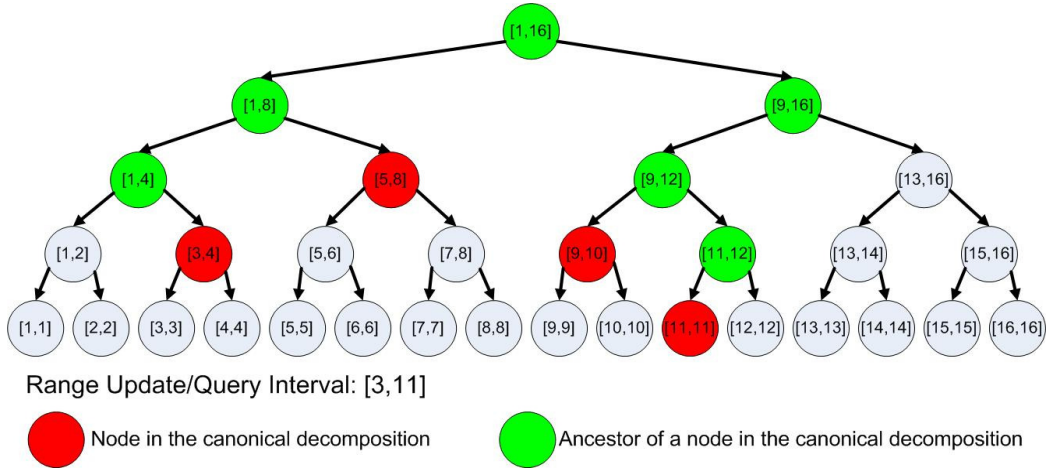

Fig. 1. A 1D segment tree with 16 leaves and a canonical decomposition of the range [3,11].

The function *parity(s(i))* returns *0* if *s(i)=chigh(i)*, and *1* if *s(i)=clow(i)-1*. If the term *(-1)·PSCube(s(1),…,s(d))* appears in the « sum », then we need to consider the *inverse* of *PSCube(s(1),…,s(d))* (the inverse of a value *x* is : *–x*, for *aggf=+* ; *1/x*, for *aggf=\** ; *x*, for *aggf=xor* ; and so on). The large « sum » symbol denotes the aggregation of the terms (or their inverses) of *PSCube*. The small « sum » symbol denotes addition. If we consider the number of dimensions *d* to be constant, then a range aggregate query can be answered in *O(1)* time. We will now show how to compute efficiently the entries of the prefix « sum » cube. We consider the sequences of coordinates *(c(1), …, c(d))* in lexicographic order and compute *PSCube(c(1), …, c(d))* as follows. If any of the coordinates is *0*, then the entry is equal to the neutral element (depending on *aggf*). Otherwise, we have :

$$PSCube(c(1),...,c(d)) = aggf(Cube(c(1),...,c(d)), \sum_{\substack{s(i) \in \{c(i)-1, c(i)\}, 1 \leq i \leq d \\ (\exists) 1 \leq j \leq d . s(j) \neq c(j)}} (-1)^{1 + \sum_{j=1}^{d} parity'(s(i))} \cdot PSCube(s(1),..., s(d))) \quad (2)$$

The function *parity'(s(i))* returns *0* if *s(i)=c(i)*, and *1*, if *s(i)=c(i)-1*. This way, every entry *PSCube(c(1), …, c(d))* is computed in $O(2^d)$ time. The large « sum » symbol and the terms *(-1)·PSCube(c(1),…,c(d))* have the same meaning as before. We can do a little better, though, as was observed in [1]. We initialize *PSCube(c(1), ..., c(d))* to *Cube(c(1), ..., c(d))* (for all the tuples *(c(1), ..., c(d))*). Then, for each dimension *i* ($1 \leq i \leq d$), in order, we compute the prefix « sums » along that dimension, i.e., in lexicographic order of the sequences of coordinates *(c(1), ..., c(d))*, we set *PSCube(c(1), ..., c(d))=aggf(PSCube(c(1), ..., c(d)), PSCube(c(1), ..., c(i-1), c(i)-1, c(i+1), ..., c(d)))*. This takes *O(d)* time per entry.

An application of the prefix « sum » technique is provided by the batched range update problem, which was briefly mentioned in [7]. Let's consider the same d-dimensional data cube as before, whose entries initially contain the neutral element. We are given a list of *q* updates. An update consists of a d-dimensional range *[xa(1),xb(1)] x ... x [xa(d),xb(d)]* ($1 \leq xa(i) \leq xb(i) \leq m(i)$, $1 \leq i \leq d$, *xa(i)* and *xb(i)* are the indices of the *xa(i)$^{th}$* and the *xb(i)$^{th}$* distinct coordinate values in dimension *i*) and an update value *u*. The effect of the update is to set each data entry *Cube(c(1), ..., c(d))* with $xa(i) \leq c(i) \leq xb(i)$ ($1 \leq i \leq d$) to *aggf(u, Cube(c(1), ..., c(d)))*. Of course, we want to perform the updates in an efficient manner, i.e. without changing the value of each data entry independently (such an approach would have a time complexity of *O(q·Np)*, where *Np=m(1)·...·m(d)*). An efficient technique is the following. For each update, we modify the $2^d$ entries *Cube(c(1), ..., c(d))*, with *c(i)* in *{xa(i), xb(i)+1}* in the following way. If the number of coordinates *c(i)* ($1 \leq i \leq d$) such that *c(i)=xb(i)+1* is even, then we set *Cube(c(1), ..., c(d))* to *aggf(u, Cube(c(1), ..., c(d)))* ; otherwise, we set it to *aggf(u$^{-1}$, Cube(c(1), ..., c(d)))*. For instance, if *aggf=+*, in the first case we increase *Cube(c(1), ..., c(d))* by *u* and in the second case we decrease it by *u* (it's similar for the other functions, e.g. multiplication and division, xor and xor, and others). In the end we compute the prefix « sum » cube *PSCube*, as shown before. *PSCube(c(1), ..., c(d))* will be equal to the final value of the entry *Cube(c(1), ..., c(d))*, after applying all the updates. The complexity is only *O((q+Np)·$2^d$)*, or *O(q·$2^d$+Np·d)*.

The techniques developed for range searching are also useful for solving problems where some unexpected transformation reduced them to a range searching problem. For instance, let's consider the following problem. We are given a rooted tree with *n* vertices. Each vertex *i* has a weight *w(i)* and every edge *(u,v)* has a length *length(u,v)*. We want to answer queries of the following form: compute the aggregate of the weights of all the vertices located in the subtree of vertex *i* and located at distance at least $d_1$ and at most $d_2$ ($d_2 \geq d_1 \geq 0$) from *i*. The subtree of a vertex *i* is composed of vertex *i* and the subtrees of its sons. We will perform a DFS traversal of the tree starting from the root. During this traversal, we assign to each vertex *i* its DFS number (*DFSnum(i)*). We have *DFSnum(i)=j* if vertex *i* was the *j$^{th}$* distinct vertex visited during the DFS traversal. All the DFS numbers of the vertices *p* in vertex *i*'s subtree are $\geq DFSnum(i)$ and they form an interval of consecutive numbers *[DFSnum(i), DFSmax(i)]*, where *DFSmax(i)* is the largest DFS number of a vertex in vertex *i*'s subtree. We also compute for each vertex $i$ the value $droot(i)$=the distance from the root to vertex $i$. We have $droot(root)=0$ and $droot(i \neq root)=length(parent(i),i)+droot(parent(i))$. We will assign to each vertex $i$ a point in the plane with coordinates $(DFSnum(i), droot(i))$ and weight $w(i)$. We insert all these points in a 2D range tree $RT$. The types of queries that we mentioned are now equivalent to range querying the range $[DFSnum(i), DFSmax(i)] \times [droot(i)+d_1, droot(i)+d_2]$ and obtaining the aggregate value of the weights of the points in this range. Thus, each query can be answered in $O(log^2(n))$ time, or $O(log(n))$ if we use the fractional cascading method. This method assumes that the range tree nodes $q$ at $d=2$ do not store a tree $T(q)$, but rather an array $A(q)$ of the points in the corresponding range (the same points that would have been stored in $T(q)$), sorted according to the points' first dimension coordinates (if $q$ is not a leaf, $A(q)$ can be obtained by merging $A(leftson(q))$ and $A(rightson(q))$). If the aggregate function $aggf$ is invertible, we compute an array of prefix aggregates at each such node $q$: $pagg(q,0)$=the neutral element and $pagg(q,i \geq 1)=pagg(q,i-1)\ aggf\ w(A(q,i))$ (where $A(q,i)$ is the $i^{th}$ point in the sorted array of node $q$). A (multidimensional) range query will ask every node $q$ at $d=2$ to run a 1D range query on an interval $[u,v]$ of positions from $A(q)$: such a range query can be answered in constant time, as $pagg(q,v)\ aggf\ (pagg(q,u-1))^{-1}$. We can also answer range maximum or minimum queries in $O(1)$ time at a node $q$, as it is well known that we can preprocess the weights of the points stored at every node $q$ (at $d=2$) in time proportional to their number, in order to answer such queries in $O(1)$ time. Usually, the indices $u$ and $v$ are computed by using a binary search (because the 1D range is $[xa(1),xb(1)]$ and we need to find the smallest position $u$ s.t. $x(A(q,u),1) \geq xa(1)$ and the largest position $v$ s.t. $x(A(q,v),1) \leq xb(1)$). The fractional cascading technique allows us to perform a binary search only at the root of the tree $T(q')$ in which a node $q$ is contained. Then, the indices $u$ and $v$ for each relevant node $q$ from $T(q')$ are computed in $O(1)$ time.

A 1D application of some of the range update and aggregation techniques we presented earlier is the following. We are given $n$ communication stations, interconnected by a network architecture similar to a bus. The stations are arranged in a line, such that station $i$ can transmit data only to station $i+1$ ($1 \leq i \leq n-1$). Each station has a data sending rate $s(i)>0$ and a maximum data processing rate $r(i)>s(i)$. In a normal state, each station $i$ sends data at a rate $s(i)$ and receives data at a rate $rr(i) \leq r(i)-s(i)$. If, however, $rr(i)>r(i)-s(i)$, the station collapses naturally and does not consume the received data anymore. As a consequence, data is transmitted at a rate of $rr(i)+s(i)$ to the next station $(i+1)$. Each station also has a security cost $c(i)$, representing the amount of effort required to make the station collapse *artificially* (i.e. set $r(i)$ to $0$). For security reasons, we want to assess the minimum total amount of effort required to make the $n^{th}$ station collapse. Note that by collapsing a subset of stations, other stations might collapse without any extra effort, due to the data receiving rate being larger than their maximum allowed rate. We will compute the prefix sums $ps(i)=s(1)+\ldots+s(i)$ ($ps(0)=0$ and $ps(i)=ps(i-1)+s(i)$). Then, using these prefix sums, for each station $i$ we will compute the smallest index $prev(i)$ ($1 \leq prev(i) \leq i$), such that $ps(i)-ps(prev(i)-1) \leq r(i)$ (we can use binary search to compute $prev(i)$). After this, for each station $i$, we will compute $e(i)$=the total effort required to collapse the $n^{th}$ station if we artificially collapse station $i$ and we do not artificially collapse any other station $j<i$. We will maintain a segment tree over the $e(i)$ values of the $n$ stations, which are initialized to $0$. Then, for each station $i$, we update the interval $[prev(i),i]$ with the value $c(i)$. This means that the value $e(j)$ of every station $j$ ($prev(i) \leq j \leq i$) should be increased by $c(i)$. By using a segment tree, we find the canonical decomposition of the interval $[prev(i),i]$ and increase the $uagg$ values of the tree nodes in the canonical decomposition. At the end, we compute the actual values $e(i)$. $e(i)$ is the sum of the $uagg(q)$ values of the tree nodes $q$ on the path between the leaf corresponding to station $i$ and the tree root (including the endpoints). The time complexity is $O(n \cdot log(n))$. Note that all the updates take place before the point queries. Thus, we could also use the technique regarding batched range updates, presented earlier, for computing all the $e(*)$ values in $O(n)$ time (the total time complexity remains $O(n \cdot log(n))$, due to the binary searches).

### 3. Selection and Range Optimization Problems

In this section we will present novel algorithmic results for several selection and range optimization problems. In the first problem we consider $n$ (ascendingly) sorted sequences of numbers. Each sequence $i$ ($1 \leq i \leq n$) has $b(i)$ distinct elements. We want to find the $k^{th}$ smallest element of the sequence obtained by merging the elements of the $n$ sequences in increasing order. However, the numbers in the $n$ sorted sequences are not known in advance. We can use a query operation $Qu(i,j)$ which returns the $j^{th}$ number ($1 \leq j \leq b(i)$) from the $i^{th}$ sequence. We would like to use the query operation as few times as possible. We will present a solution which does not necessarily perform the minimum number of queries, but which is, nevertheless, quite efficient. We will maintain for every sequence $i$, the potential interval of positions $[low(i),high(i)]$ in which the number we are looking for may reside – initially, $low(i)=1$ and $high(i)=b(i)$. The meaning of $low(i)$ ($high(i)$) is that we know for sure that all the numbers in the sequence $i$ located on positions which are strictly smaller (larger) than $low(i)$ ($high(i)$) are strictly smaller (larger) than the element we are looking for. We will perform several iterations, as follows. In every iteration, we will choose any sequence $q$, such that $low(q)<high(q)$. Then, we will set $mid(q)=(low(q)+high(q))$ $div\ 2$. Moreover, for every sequence $i$ we will maintain a data structure $DS(i)$ (e.g. a balanced tree) with values $j$ and associated values $x(i,j)$ (i.e. the value on position $j$ of the sequence $i$), for all the positions $j$ for which the values are known in the sequence $i$. We will try to find the value $x(q,mid(q))$. If the position $q$ is in $DS(q)$, then we already know the value; otherwise, we perform a query $Qu(q,mid(q))$ and we find out $x(q,mid(q))$ (then, we insert $mid(q)$ with the associated value $x(q,mid(q))$ into $DS(q)$). After this, we will compute $nv(i)$ for every sequence $i$: $nv(i)$=the number of values in the sequence $i$ which are smaller than or equal to $x(q,mid(q))$. Obviously, $nv(q)=mid(q)$. For every other sequence $i$, we will find the two positions $u(i)$ and $v(i)$ from $DS(i)$ with known values on them, such that $x(i,u(i)) \leq x(q,mid(q))$ (and $u(i)$ is the largest with this property) and $x(i,v(i))>x(q,mid(q))$ (and $v(i)$ is the smallest with this property); we also consider that $DS(i)$ contains the fictitious positions $0$, with $x(i,0)=-\infty$, and $b(i)+1$, with $x(i,b(i)+1)=+\infty$. Then, we will perform a binary search on the interval $[u(i),v(i)-1]$. Let $ulow=u(i)$, $uhigh=v(i)-1$ and $uok=u(i)$. While $ulow \leq uhigh$, we perform the following steps: *(1)* we set $umid=(ulow+uhigh)$ *div* $2$; *(2)* if $x(i,umid)$ is not known, we query $Qu(i,umid)$ (and then we insert the values $umid$ and the associated value $x(i,umid)$ in $DS(i)$); *(3)* if $x(i,umid) \leq x(q,mid(q))$ then we set $uok=umid$ and $ulow=umid+1$; otherwise, we set $uhigh=umid-1$. At the end of this binary search, we set $nv(i)=mid(i)=uok$. Then, we compute the sum of the $nv(*)$ values: $snv=nv(1)+\ldots+nv(n)$. If $snv<k$ then, for every sequence $i$, we set $low(i)=max\{low(i), mid(i)+1\}$; if $snv>k$ then, for every sequence $i$, we set $high(i)=min\{high(i), mid(i)\}$. If $snv=k$ then the $k^{th}$ smallest value in all the sequences is $x(q,mid(q))$ and the algorithm ends. The iterative process also stops when there are no more sequences $i$ with $low(i)<high(i)$ (or, alternatively, after we obtain $snv>k$ and $snv-k<n$ and we update the $high(*)$ values accordingly). Then, if the $k^{th}$ smallest value has not been found, yet, we compute $snv$=the sum of the $high(i)$ values of all the sequences: $snv=high(1)+\ldots+high(n)$. Then, since the $k^{th}$ smallest element has not been found, yet, we must have $snv>k$. However, we have that $snv-k<n$. We initialize $idx(i)=high(i)$ for every sequence $i$. Then, we insert in a max-heap $H$ all the values $x(i,idx(i))$, together with their associated sequence index $i$ (with $1 \leq i \leq n$ and $idx(i)>0$); if the value is not known, we perform a query for it. While $snv>k$ : *1)* we extract the largest value $x(i,idx(i))$ (corresponding to a sequence $i$) from $H$; *2)* we decrease $idx(i)$ by $1$; *3)* if $idx(i)>0$ we insert $x(i,idx(i))$ into $H$ (together with the associated sequence index $i$) – if $x(i,idx(i))$ is not known, we perform a query for it; *4)* we decrement $snv$ by $1$. In the end, when $snv=k$, the largest value in $H$ is the $k^{th}$ smallest value we were searching for. The algorithm performs at most $n \cdot log(max\{b(i)|1 \leq i \leq n\})$ queries per iteration, and the number of iterations can be $O(n \cdot log(max\{b(i)|1 \leq i \leq n\}))$. In the end, the algorithm may perform $O(n)$ extra queries. This algorithm can be extended to finding the $k^{th}$ smallest value among the numbers on the positions $[a(i),b(i)]$ of every sequence $i$: we just replace every sequence $i$ by a sequence $i'$ consisting of the $b(i)-a(i)+1$ positions $[a(i),b(i)]$ of the sequence $i$, and we translate $Qu(i',j)$ into $Qu(i,j+a(i)-1)$.

As the second problem we consider the multidimensional dynamic range

Practical Range Aggregation, Selection and Set Maintenance Techniques

median problem. We have a d-dimensional hyper-cube, with $m(j)$ cells in each dimension $j$ ($1 \leq j \leq d$). The $k^{th}$ cell in the $j^{th}$ dimension has an assigned coordinate: $x(j,k)$ ($x(j,k) \leq x(j,k+1)$ for $1 \leq k \leq m(j)-1$). Each cell $(c(1), ..., c(d))$ has a value $Cube(c(1), ..., c(d))$. Given a d-dimensional range *[clow(1), chigh(1)] x ... x [clow(d), chigh(d)]* (with $1 \leq clow(j) \leq chigh(j) \leq m(j)$ for every $1 \leq j \leq d$), we want to find the location of a point *P* such that the sum of weighted $L_1$ distances from *P* to every point assigned to a cell in the range is minimum (the point assigned to a cell $(c(1), ..., c(d))$ is $(x(1,c(1)), ..., x(d,c(d)))$); the weighted $L_1$ distance to the point corresponding to a cell *(c(1), ..., c(d))* is equal to ($Cube(c(1), ..., c(d))$ multiplied by the actual $L_1$ distance). In [4] an efficient solution to the static version of this problem was given. Here we augment that solution by allowing point updates (i.e. the value of a cell can be modified) and restricted range updates (i.e. the value of each point in a given range is updated, e.g. it is increased, by the same value *u*); our time complexities are: $O(d \cdot log^d(n))$ for a point update and $O(n \cdot d \cdot log^d(n))$ for a range update. The technique in [4] is based on being able to compute efficiently the sum of the *XCube(*,...,*)* values in any given d-dimensional range of a d-dimensional array *XCube*. When the hyper-cube is static, this can be done in $O(2^d)=O(1)$ time. If we construct a d-dimensional segment tree over the cells of a hyper-cube *H*, then point updates, some restricted range updates, and range sum queries can be performed in $O(log^d(n))$ time each over the cells of *H* ($n=max\{m(j) \mid 1 \leq j \leq d\}$); see [7]. Thus, the query time complexity of the method given in [4] is increased by an $O(log^d(n))$ factor in this version of the problem. When a range *[clow(1), chigh(1)] x ... x [clow(d), chigh(d)]* (possibly just one cell) of *Cube(c(1), ..., c(d))* is modified by *v* (e.g. each value in the range is increased by *v*), we need to update the range in the *Cube* d-dimensional array by *v*. Then, we need to (range) update every range *[clow(1), chigh(1)] x ... x [clow(j-1), chigh(j-1)] x [c(j), c(j)] x [clow(j+1), chigh(j+1)] x ... x [clow(d), chigh(d)]* in the *DCube$_j$* d-dimensional array by $x(j,c(j)) \cdot v$ (for every cell $clow(j) \leq c(j) \leq chigh(j)$; $1 \leq j \leq d$).

    A related median finding problem is the following. Given *n* points on the real line (with point *i* at coordinate $x(i)$; $1 \leq i \leq n$), we want to find the location of a point *xp* such that the sum of distances from each of the *n* points to *xp* is minimum. We consider two cases. In case *1*, the distance from point *i* to *xp* is $|x(i)-xp|$. In case *2*, every point *i* also has a non-negative weight $w(i)$ and the distance is defined as $w(i) \cdot (x(i)-xp)^2$. For case *1*, let's consider the points sorted such that $x(1) \leq .. \leq x(n)$. *xp* can be located anywhere inside the interval *[x(1),x(n)]*, without changing the sum of distances to the points *1* and *n*. In a similar manner, *xp* can be located anywhere inside *[x(2), x(n-1)]*, and so on. If *n* is odd then $xp=x((n+1)/2)$. If *n* is even, then *xp* can be any point inside the interval *[x(n/2), x((n/2)+1)]* (even $xp=x(n/2)$). Thus, in order to compute *xp*, we can sort all the *n* values in $O(n \cdot log(n))$ time, or we can use a linear time algorithm for selecting the median value of the (multi-)set of numbers $x(1), ..., x(n)$.

In case *2* we must minimize the sum $S(xp)=w(1) \cdot (x(1)-xp)^2+...+w(n) \cdot (x(n)-xp)^2$. The derivative of *S*, *dS/dxp* must be equal to *0*. Thus, we must have $2 \cdot (w(1) \cdot x(1)+...+w(n) \cdot x(n))=2 \cdot (w(1)+...+w(n)) \cdot xp$ => $xp=(w(1) \cdot x(1)+...+w(n) \cdot x(n))/(w(1)+...+w(n))$ (i.e. *xp* is the weighted average of the *n* x-coordinates). In this case, the solution also has a linear time complexity.

For the third problem we consider a sequence of *n* numbers: *S(1), ..., S(n)* and a sequence of *m* operations of three types: type 1) **R(i,j)** ($1 \leq i \leq j \leq n$) reverses the order of the numbers on the positions *i, ..., j* ; type 2) **C(i,j,p)** ($1 \leq i \leq j \leq n$; $-1 \leq p \leq n-(j-i+1)$) *cuts* the numbers *S(i), ..., S(j)* from the sequence and *pastes* them, in the same order, after the position *p* of the remaining sequence (if *p=-1* then no paste occurs) ; type 3) **I(p, k, $v_1$, ..., $v_k$)** – inserts the numbers $v_1$, ..., $v_k$ after the position *p* of the sequence ; type 4) **Q(i)** asks for the current value of *S(i)* (the current number on the position *i* in the sequence). We will start by presenting a solution which works well when *m* is not too large. We will maintain a sequence *SI* of *([a(u),b(u)], dir(u))* pairs, where the intervals correspond to the order in which the elements occur in the sequence. Such an interval *[a(u),b(u)]* occurring on the position *u* of *SI* will have the meaning that the numbers *So(a(u)), ..., So(b(u))* from the original sequence *So* are located on consecutive positions in the current sequence. If *dir(u)=+1* then these numbers occur in increasing order of the positions *a(u), ..., b(u)*; if *dir(u)=-1* then they occur in reverse order. Initially, we only have one interval, *[a(1)=1,b(1)=n]*, and *dir(1)=+1*. Then, we traverse the sequence of operations (in order). All the operations will make use of the following function: *Find(i)*. If *i=0* then *Find(i)* returns *0*. Otherwise, *Find(i)* works as follows: We will traverse the *(interval, direction)* pairs from *SI* from left to right (i.e. starting from the first position) and we will maintain a counter *k*, representing the number of positions traversed so far (initially *k=0*). When we reach a position *p* in *SI*, we increase *k* by *(b(p)-a(p)+1)*. If, after considering the $u^{th}$ interval in *SI*, we have *k≥i*, then *[a(u),b(u)]* is the interval containing the position *i*. Let *q=i-(k-(b(u)-a(u)+1))* be the position in the interval corresponding to position *i*. If *dir(u)=+1* then we will split the interval *[a(u),b(u)]* into *x=*(at most) *3* intervals: *[a(u),a(u)+q-2]*, *[q'=a(u)+q-1,q'=a(u)+q-1]* and *[a(u)+q,b(u)]* (we disregard any empty intervals among these three). We will insert these intervals (in this order) instead of the interval on position *u* of *SI*, setting their corresponding *dir* values to *+1* (not before shifting *x-1* positions to the right every interval and *dir* value from a position larger than *u*). If *dir(u)=-1*, then we split the interval into *x=*(at most 3) intervals: *[b(u)-q+2,b(u)]*, *[q'=b(u)-q+1,q'=b(u)-q+1]* and *[a(u),b(u)-q]*. Like before, we insert the non-empty intervals among these intervals in *SI* (in this order) instead of the former interval on position *u* and we set their *dir* values to *-1* (we also shift the other *(interval, direction)* pairs to the right, like before). *Find(i)* returns *h*, such that *a(h)=b(h)=q'*.

At an operation *R(i,j)* we will first compute *u'=Find(i)* and then *v'=*

Practical Range Aggregation, Selection and Set Maintenance Techniques

*Find(j)*. Then, we will reverse the order of the *(interval, direction)* pairs between the positions *u'* and *v'* from the *SI* array. Basically, we swap the *(interval, direction)* pairs from the positions *po* and *(v'-(po-u'))* (for *u'≤po≤(v'+u') div 2*). Afterwards, we set *dir(po)=-dir(po)* for every position *u'≤po≤v'*. In the case of a *C(i,j,p)* operation, we compute *u'=Find(i)* and then *v'=Find(j)*. Then, we construct *SI'* by removing from *SI* all the *(interval, direction)* pairs on the positions between *u'* and *v'* inclusive. Then, we compute *w'=Find(p)* (using the *SI'* array). Then, we insert the *(interval, direction)* pairs removed previously (their *a*, *b*, and *dir* values) in *SI'*, right after the position *w'* and we set *SI* to the obtained sequence of *(interval, direction)* pairs. If, however, we have *p=-1* then we set *SI=SI'*. An *I(p, k, $v_1$, …, $v_k$)* operation is handled as follows. We add the numbers $v_1$, …, $v_k$ at the end of the original sequence *So*: *(So(n+i)=$v_i$; 1≤i≤k)* and then we set *n=n+k*. Then, we compute *w'=Find(p)*. After the position *w'*, we insert in the array *SI* the pair *([n-k+1,n],1)*. In order to answer a query *Q(i)*, we find the interval *[a(u),b(u)]* containing the position *i* (by using the same counter *k* as in the function *Find*) and the value of *q=i-(k-(b(u)-a(u)+1))*. This time, however, we will not split the interval *u*. If *dir(u)=+1*, then the answer is *So(a(u)+q-1)*; if *dir(u)=-1*, then the answer is *So(b(u)-q+1)*. Here, *So(p)* denotes the number on the position *p* in the initial sequence (on which no changes were performed). The time complexity is *O(min{m,n})* for each *R, C* or *Q* operation, and *O(min{m,n}+k)* for each *I* operation (where *k* is the number of inserted elements). The overall time complexity is *O(min{m·n,$m^2$}+n)*, where *n* is the total number of elements (the initial elements plus the newly inserted elements).

    In order to solve the problem more efficiently we make use of an idea mentioned to us by M. Paşoi. We will split the sequence of *m* operations into groups of *z* operations (the last group may contain fewer than *z* operations). For each group of *z* operations, we will use the methods described above. After processing all the *z* operations in the $g^{th}$ group *(1≤g≤ng; ng=O(m/z)* is the total number of groups), we will generate the contents of the sequence *S* of numbers at that moment in time, in *O(n+min{n,z})* time. We will traverse the sequence of intervals *SI* from position *1* to *|SI|* (*|SI|*=the total number of intervals in *SI*) and, at the same time, we will maintain a counter *k* (initially, *k=0*). When we reach the $u^{th}$ interval, if *dir(u)=+1*, then we set *S(k+j)=So(g-1, a(u)+j-1) (1≤j≤b(u)-a(u)+1)*; if *dir(u)=-1*, then we set *S(k+j)=So(g-1, b(u)-j+1) (1≤j≤b(u)-a(u)+1)*. Then, we increment *k* by *(b(u)-a(u)+1)*. Here, *So(g-1)* denotes the sequence right before performing the first operation from the current group. If *g=1*, then *So(0)* is the original sequence. After computing *S*, we set *So(g)=S* and *n=k*. Note that after computing *So(g)*, we do not need the previous sequences *So(0≤g'≤g-1)*, which can be dropped (in order to use only *O(n)* memory). The time complexity of this approach is *O(m/z·min{n·z,n+$z^2$})=O(min{m·n/z+m·z,m·n})*. We can choose, for instance, *z=O(sqrt(n))* or *z=O(sqrt(m))*, obtaining an *O(m·sqrt(n))* or an

$O(min\{(n+m) \cdot sqrt(m), m \cdot n\})$ time complexity, which is much better than $O(min\{m \cdot n, m^2\}+n)$; $sqrt(h)$ denotes the square root of $h$.

A related problem is the following. We consider a stack (initially empty), on which we perform $M$ operations of the following two types: *1) Push(x)*: push an element $x$ at the top of the stack; *2)* rotate the topmost $K$ elements of the stack ($K$ is the same for each such operation). We want to print the final order of the elements in the stack (from bottom to top). In order to solve the problem we will construct a stack $F$ (initially empty), containing the final order of the elements. We will also construct an array $V$ which is sufficiently large (e.g. it has at least $2 \cdot M$ elements) and we will maintain two pointers, *up* and *down* (*up* points to the topmost element and down points to the $K^{th}$ element from the top, or the bottom element if there are less than $K$ elements in the stack). We will also maintain the direction *dir* in which the next element will be added. We will start with *down=M*, *up=down-1* and *dir=+1*. At every *Push(x)* operation, we set *up=up+dir* and then *V(up)=x*. If at least $K+1$ *Push* operations have been performed so far (including the current one), then we push *V(down)* at the top of $F$ and then we set *down= down+dir*. At a rotation operation, we swap the values of *up* and *down* and set *dir=-dir*. After performing all the operations, we traverse all the elements between *down* and *up* from the array $V$ (*V(down)*, *V(down+dir)*, *V(down+2·dir)*, …, *V(up)*) and we push each of them on top of $F$. Thus, each operation takes $O(1)$ time. Instead of the array $V$ we could use a doubly-linked list in which every node has two neighbors, corresponding to the directions *-1* and *+1* (then, we would replace *x=x+dir* by *x=x->neighbor(dir)*, where *x=up* or *down*; in the end, we would traverse the list from *down* and following the direction *dir* until we reach *up*).

The following interesting geometric selection problem was mentioned to us by R. Berinde in a private communication. We have $n$ points in the plane (point $i$ lies at coordinates $(x(i),y(i))$ and $y(i) \geq 0$). We have $m$ queries of the following type: what is the $k_j^{th}$ smallest (Euclidean) distance from a point $i$ with $x(i) \leq xq_j$ to the point $(xq_j, 0)$ ? The problem is offline, meaning that all the queries are known in advance. For each query $j$, we could sort all the $n$ points according to their distance to $(xq_j, 0)$ and then just select the $k_j^{th}$ distance. This approach has an $O(m \cdot n \cdot log(n))$ time complexity. An improvement consists of selecting the $k_j^{th}$ smallest distance without sorting the distances for each query, by using the *QuickSelect* selection algorithm [8] (inspired from the *QuickSort* algorithm). This approach has an $O(m \cdot n)$ time complexity. In order to obtain a better time complexity, we will proceed as follows. We will sort the coordinates $xq_j$ of the queries, such that we have $xq_{p(1)} \leq xq_{p(2)} \leq \ldots \leq xq_{p(m)}$. We will sort the points $i$ with $x(i) \leq xq_{p(1)}$ in increasing order of their distance from $(xq_{p(1)}, 0)$. Thus, we have an order of the points: $od(1)$, …, $od(np(1))$ ($np(1)$ is the number of points $i$ with $x(i) \leq xq_{p(1)}$). We will sweep the plane from left to right with a vertical line. At every moment, we will want to maintain the sorted order of the points $i$ with coordinates $x(i) \leq xd$, according to their distance from $(xd,0)$, where $x=xd$ is the current position of the sweep line. As we sweep the line, we have *3* types of events: *1)* insertion of a new point *i*: a new point is inserted when we reach $xd=x(i)$; *2)* query – such an event occurs when $xd=xq_j$ for some *j* ($1 \leq j \leq m$); *3)* swapping the order of two points located on consecutive positions in the ordering according to distance. Event *3* is based on the following observation. Let's assume that we have the points *od(i)* and *od(i+1)*, in increasing order of their distance from *(xd,0)*. If $x(od(i))<x(od(i+1))$ (the *x-condition*) then there exists a coordinate *xsod(od(i))*, up until which *od(i)* will be located before *od(i+1)* in the distance ordering, and after which the order of *od(i)* and *od(i+1)* will be swapped – from there on, *od(i+1)* will always be located before *od(i)* in the ordering. We can compute this coordinate from the equation: $(x(od(i))-xsod(od(i)))^2+y(od(i))^2 = (x(od(i+1))-xsod(od(i)))^2+y(od(i+1))^2$. This equation is, in fact, only a first degree equation with the unknown *xsod(od(i))*. Thus, we can use the following solution. We will maintain a heap with events of type *3* and two arrays of events of types *1* and *2*. At every step, we will choose the event located at the smaller coordinate (either the next event of type *1* or *2* from the corresponding array, or the event with the minimum x-coordinate from the heap; if multiple events of different types take place at the same x-coordinate, we give preference to type *1* events first). Initially, we compute the values *xsod(od(i))* for the initial order of the points (relative to the distance to $(xq_{p(1)},0)$) and only for those points for which the *x-condition* is met; we insert the values *xsod(od(i))* (together with the values *od(i)* and *od(i+1)*) in *H* ($1 \leq i \leq np(1)$). For a type *1* event, we will search the position *p* on which the point has to be inserted in the distance ordering (distances are computed relative to *(xd,0)*, where *xd* is the x-coordinate of the event). If we insert the new point on the position *i* (the new point is *od(i)*), then we compute the values *xsod(od(i-1))* (if *i>1*) and *xsod(od(i))* (if *i* is not the currently last position) – only if the corresponding *x-conditions* are met - and insert them into *H* (together with the indices of the two points to which the event corresponds). We can store the points *od(i)* in an array, in which case an operation of type *1* can be handled in $O(log(n))$ (for binary searching the position *i*) + $O(n)$ (for shifting to the right the points located after the position *i* in the *od* array) time. A type *2* event is handled in $O(1)$ time, by setting the distance from $od(k_{p(i)})$ to $(xq_{p(i)},0)$ as the answer to the query *p(i)* to which the type *2* event corresponds. A type *3* event, corresponding to the swap of the points *a* and *b*, is handled as follows. For each point *i* ($1 \leq i \leq n$) we will maintain a value *pos(i)=p*, if *od(p)=i* (or *pos(i)=0* if *i* was not inserted in the *od* array, yet). Whenever we insert a new point, we shift a point to the right, or we swap the order of two points in the *od* array, we also update the *pos* values of the corresponding point(s). Let's assume that *pos(a)<pos(b)* and *pos(b)=pos(a)+1*. Then, we swap *od(pos(a))* with *od(pos(b))*. Before performing the swap, we delete from *H* the events corresponding to the swap of *a* with *od(pos(a)-1)* (*xsod(od(pos(a)-1))*) and to the swap of *b* with *od(pos(b)+1)* (*xsod(b)*), if such events exist in *H*. After performing the swap (and adjusting the *pos* values accordingly), if *pos(a)* is not the last position in the *od* array (*pos(b)>1*), we insert in *H* a type *3* event with *x=xsod(od(pos(a)))* (*x=xsod(od(pos(b)-1))*) corresponding to the swap of the points *od(pos(a))* (*od(pos(b))*) and *od(pos(a)+1)* (*od(pos(b)-1)*); the events are considered only if the corresponding *x-conditions* are met. A type *3* event is handled in $O(log(n))$ time. Overall, there are *n* type *1* events, *m* type *2* events (which need to be sorted) and $O(n^2)$ type *3* events. The time complexity is $O(n^2+m \cdot log(m)+n^2 \cdot log(n))=O(m \cdot log(m)+n^2 \cdot log(n))$. If the queries are given in increasing order of $xq_j$, then sorting the type *2* events is no longer required and the total time complexity becomes $O(m+n^2 \cdot log(n))$.

Another related problem (the *Longest Common Contiguous Subsequence*) is mentioned in [6]. The algorithm in [6] contains a small error, which we correct here: *(2) the first position of su(left) belongs to a string S(j) for which ((x(j)>a(j)) or ((x(j)=a(j)) and (nok>F)) or ((x(j)<a(j)) and (nok≥F)))* : in this case we set *x(j)=x(j)-1* first (and, if *x(j)* becomes equal to *a(j)-1*, we also decrement *nok* by *1*).

## 4. Conclusions

In this paper we presented several novel, practical, range aggregation and selection techniques and we discussed several important set maintenance issues.